\documentclass[conference,a4paper]{IEEEtran}
\IEEEoverridecommandlockouts
\usepackage{cite}
\usepackage{amsmath,amssymb,amsfonts}
\usepackage{algorithm}
\usepackage{algorithmic}
\usepackage{graphicx}
\usepackage{textcomp}
\usepackage{xcolor}
\usepackage{booktabs}
\usepackage{multirow}
\usepackage{array}
\newcolumntype{L}[1]{>{\raggedright\arraybackslash}m{#1}}
\newcolumntype{C}[1]{>{\centering\arraybackslash}m{#1}}
\usepackage{url}
\usepackage{placeins}
\usepackage[caption=false,font=footnotesize]{subfig}
\usepackage{dblfloatfix}
\usepackage{hyperref}
\usepackage{xurl}
\newcommand{\code}[1]{\nolinkurl{#1}}
\graphicspath{{./}}
\def\BibTeX{{\rm B\kern-.05em{\sc i\kern-.025em b}\kern-.08em
    T\kern-.1667em\lower.7ex\hbox{E}\kern-.125emX}}

\begin{document}

\title{Reproducible Validation of Voucher-Based L2 Interoperability: Diagnosing an ERC-4337 Compatibility Issue in an EIL SDK Implementation}

\author{
\IEEEauthorblockN{
Cheng-En Lee\IEEEauthorrefmark{1},
Yu-Chien Huang\IEEEauthorrefmark{1},
and Yun-Cheng Tsai\IEEEauthorrefmark{2}
}

\IEEEauthorblockA{\IEEEauthorrefmark{1}
Department of Technology Application and Human Resource Development,\\
National Taiwan Normal University, Taipei, Taiwan}

\IEEEauthorblockA{\IEEEauthorrefmark{2}
PecuLab LLC, Seattle, WA, USA\\
Corresponding author: peculab.ai@gmail.com}
}
\maketitle

\begin{abstract}
Ethereum Layer-2 (L2) ecosystems improve scalability but also fragment users, liquidity, gas funding, and execution across rollups. Consequently, cross-rollup interoperability is not only a bridging problem but also a wallet, execution, and validation problem. Ethereum Interop Layer (EIL) proposes a voucher-based architecture in which users create voucher requests on an origin chain and redeem XLP-signed vouchers on a destination chain. When reproducing the evaluated SDK version in a controlled local environment, we observed a compatibility issue in the \texttt{UserOperation} path: paymaster-related data can differ after signing, preventing a stable comparison between the user-authorized representation and the representation later inspected by the local validation flow.

This paper presents a reproducible two-L2 validation framework and a controlled compatibility mitigation for that issue. We build a deterministic local testbed over Arbitrum- and Optimism-style development chains, deploy the core paymaster and bridge-related components, implement mock bundlers and event-driven XLP providers, and introduce a sanitized paymaster-data handling path together with a compatible multichain account wrapper. Using this framework, we execute the core voucher lifecycle from request creation to destination-chain voucher redemption and asset release.

The contribution is an empirical diagnosis of an implementation-level compatibility barrier, a bounded mitigation that restores controlled end-to-end execution, and an inspectable validation artifact for studying voucher-based interoperability. The work does not claim a new interoperability protocol, universal wallet compatibility, or production readiness; it identifies the remaining gaps toward standard-account validation, one-signature multichain authorization, and full dispute-settlement support.
\end{abstract}


\section{Introduction}

Ethereum Layer-2 (L2) ecosystems increase throughput and reduce transaction cost, but they also fragment users, liquidity, gas funding, and application execution across rollups. For wallet and dApp developers, a multichain workflow must be executable, independently inspectable, and sufficiently reproducible to distinguish protocol behavior from implementation artifacts. Ethereum Interop Layer (EIL) addresses part of this challenge through an account-abstraction-based architecture in which users create voucher requests on an origin chain and redeem XLP-signed vouchers on a destination chain \cite{eil_overview,eip4337}. Because our validation environment is built over Arbitrum- and Optimism-style development chains, it also reflects representative L2 execution environments \cite{arbitrum_nitro,op_stack_interop}.

When reproducing the evaluated SDK version in a controlled local environment, we observed an implementation-level compatibility issue in the \texttt{UserOperation} handling path: paymaster-related data can differ after signing, making it difficult to preserve a stable relationship between the operation representation authorized by the user and the representation inspected by the local validation flow. This observation does not invalidate EIL's architectural design, nor does it establish a defect in every EIL implementation. Rather, it identifies a reproducibility barrier in the specific evaluated SDK configuration and motivates a carefully bounded validation harness.

We therefore build a reproducible two-L2 validation framework and introduce a controlled compatibility mitigation for the evaluated execution path. The framework provides deterministic deployment, mock bundlers, and event-driven XLP providers, enabling inspection of the core voucher-based communication and execution lifecycle. It validates the sequence from voucher request creation to destination-chain execution while explicitly separating what is demonstrated from what remains outside scope. In particular, the prototype does not establish universal ERC-4337 wallet compatibility, a production \texttt{SimpleAccount} path, one-signature multichain authorization, production bridge settlement, or dispute resolution.

The primary contribution is not a new interoperability protocol or a claim of universal standards compliance. It is a reproducible validation artifact that diagnoses an SDK-specific execution incompatibility and shows how a controlled compatibility layer can restore end-to-end testability of the voucher request--issue--redeem path. The main contributions are:
\begin{itemize}
    \item We document an implementation-level compatibility issue between post-signature paymaster-data handling and the hash-consistency assumptions required for controlled \texttt{UserOperation} validation in the evaluated SDK configuration.
    \item We design a reproducible two-L2 experimental framework with deterministic deployment, mock bundlers, and event-driven XLP providers for inspecting voucher-based interoperability workflows.
    \item We implement a bounded compatibility mitigation that enables controlled execution of the core voucher lifecycle, while clearly distinguishing it from a general SDK-level or protocol-level fix.
    \item We define the prototype boundary by identifying the executable request--voucher--redeem path, the account-abstraction assumptions used by the harness, and the remaining work needed for standard-account and production deployment validation.
\end{itemize}

\section{Background and Related Work}

\subsection{Voucher-Based L2 Interoperability}
EIL organizes cross-chain execution around a voucher lifecycle. A user first creates a voucher request on an origin chain. An off-chain provider in the interoperability layer then observes the request, evaluates the required state and policy conditions, and produces an authorization artifact that can be redeemed on a destination chain \cite{eil_overview}. This design is appealing because it separates request creation, off-chain validation, and destination-chain execution into distinct stages that can in principle be implemented across heterogeneous L2 networks.

The architecture also leverages ERC-4337-style account abstraction, in which a user action is packaged as a \texttt{UserOperation} and then validated and forwarded through a bundling flow rather than being submitted only as a traditional externally owned account transaction \cite{eip4337,erc4337_docs,erc7579}. In such a design, faithful validation of the signed operation is essential because the hash committed by the signer should remain consistent with the fields later checked by the entry-point validation logic.

\subsection{Verification Boundary in the Evaluated SDK}
Our reproduction effort revealed a compatibility issue between the intended validation semantics and the evaluated SDK execution path. Specifically, paymaster-related data may be altered after signing, causing the object later presented to the local validation flow to differ from the representation initially authorized by the user. In account-abstraction systems, this distinction is security-relevant: an evaluation harness must be precise about which bytes are committed by the authorization process and which later bytes, if any, are excluded by the applicable standard or implementation convention.

Accordingly, this work does not claim that all paymaster-data mutation is invalid or that every EIL implementation is affected. The evaluated behavior is bounded to the SDK version and local configuration listed in Table~\ref{tab:setup}. Our contribution is to diagnose how this behavior blocks clean-room, end-to-end reproduction in that configuration and to provide a controlled mitigation for research validation. The paper uses ``hash-consistency'' to mean consistency between the representation treated as authorization-relevant by the local harness and the representation recomputed during the controlled validation flow; it does not imply out-of-the-box compatibility with all public bundlers or smart-account implementations.

A second class of reproduction barriers arises from inconsistent chain-identifier handling in the evaluated SDK. Several chain-indexed maps store configuration under \texttt{BigInt(chainId)} keys but later query the same maps using non-normalized \texttt{chainId} values. We observed this pattern in the environment, builder, contract, client, and bundler-manager layers, where it can prevent resolution of paymaster addresses, entry-point addresses, chain-specific clients, and bundler endpoints even when the underlying configuration is otherwise correct. This \texttt{BigInt} inconsistency is a practical reproduction diagnosis, not a protocol-level finding.

\subsection{Related Work}
Cross-chain interoperability has been studied through message-passing bridges, settlement layers, and rollup interoperability stacks. Arbitrum Nitro and the OP Stack provide representative L2 execution environments and interoperability-oriented design points for modern Ethereum rollups \cite{arbitrum_nitro,arbitrum_overview,op_stack,op_stack_interop}. These systems, however, do not by themselves resolve wallet-level multichain authorization or voucher-based gas abstraction.

ERC-4337 introduces an account-abstraction execution model in which users authorize \texttt{UserOperation} objects that are later handled by bundlers and validated through an entry-point contract \cite{eip4337,erc4337_docs}. EIL builds on this model and extends it toward multichain execution using vouchers and XLP providers \cite{eil_overview}. The relevant literature and specifications distinguish protocol design from the correctness of concrete execution and validation paths. This paper is situated in the latter category: it evaluates whether an existing voucher-based design can be reproduced, inspected, and exercised under explicitly stated harness assumptions.

Unlike work that proposes a new bridge, settlement protocol, or account-abstraction mechanism, this work contributes a bounded compatibility study and reproducibility artifact. Its claims are limited to the evaluated SDK configuration and the controlled request--voucher--redeem workflow; it does not establish generality across other interoperability protocols, production bundlers, or wallet implementations.

\begin{figure}[!ht]
\centering
\includegraphics[width=\columnwidth]{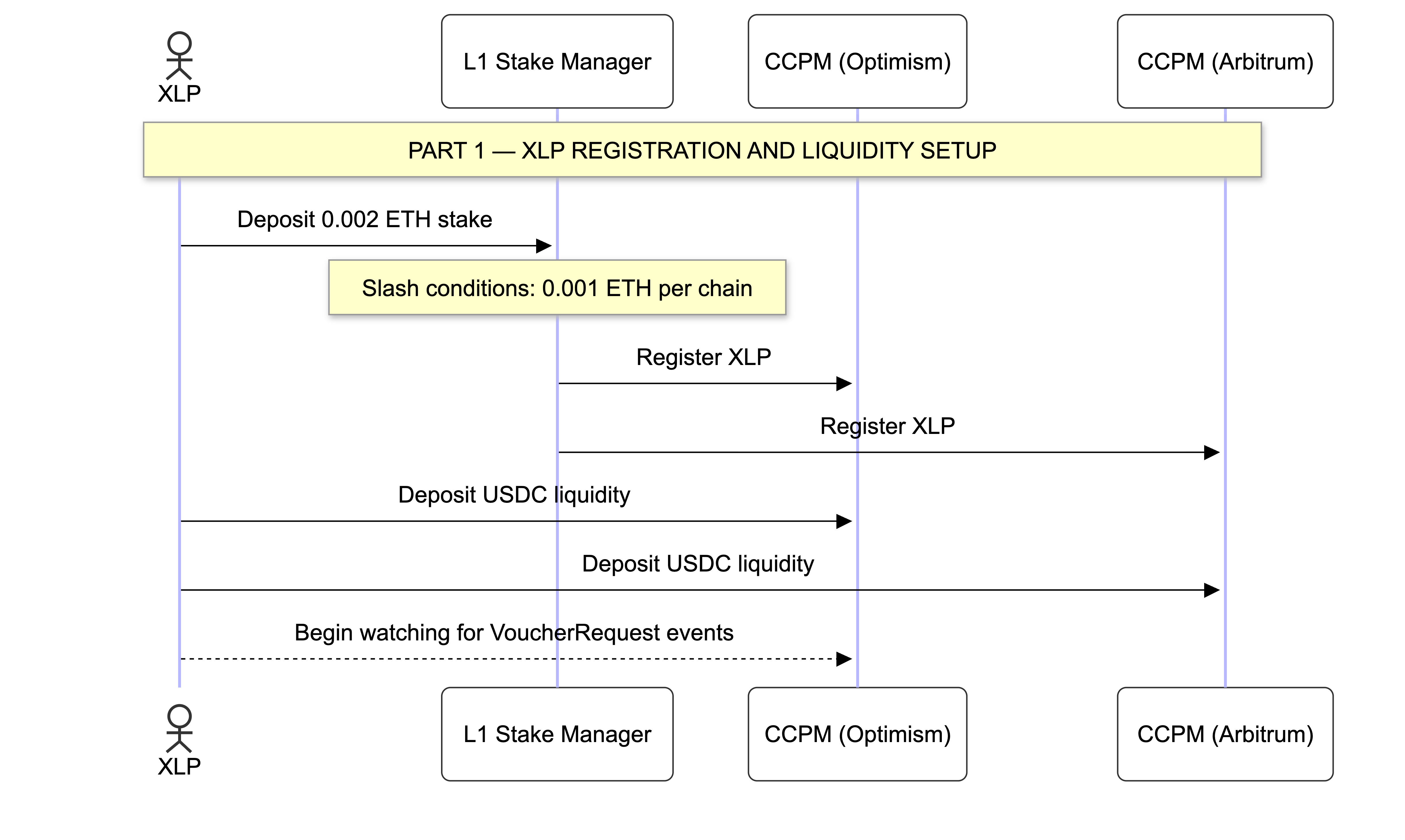}
\caption{XLP registration and liquidity setup across the two L2 environments.}
\label{fig:prep-demo-a}
\end{figure}

\begin{figure}[!ht]
\centering
\includegraphics[width=\columnwidth]{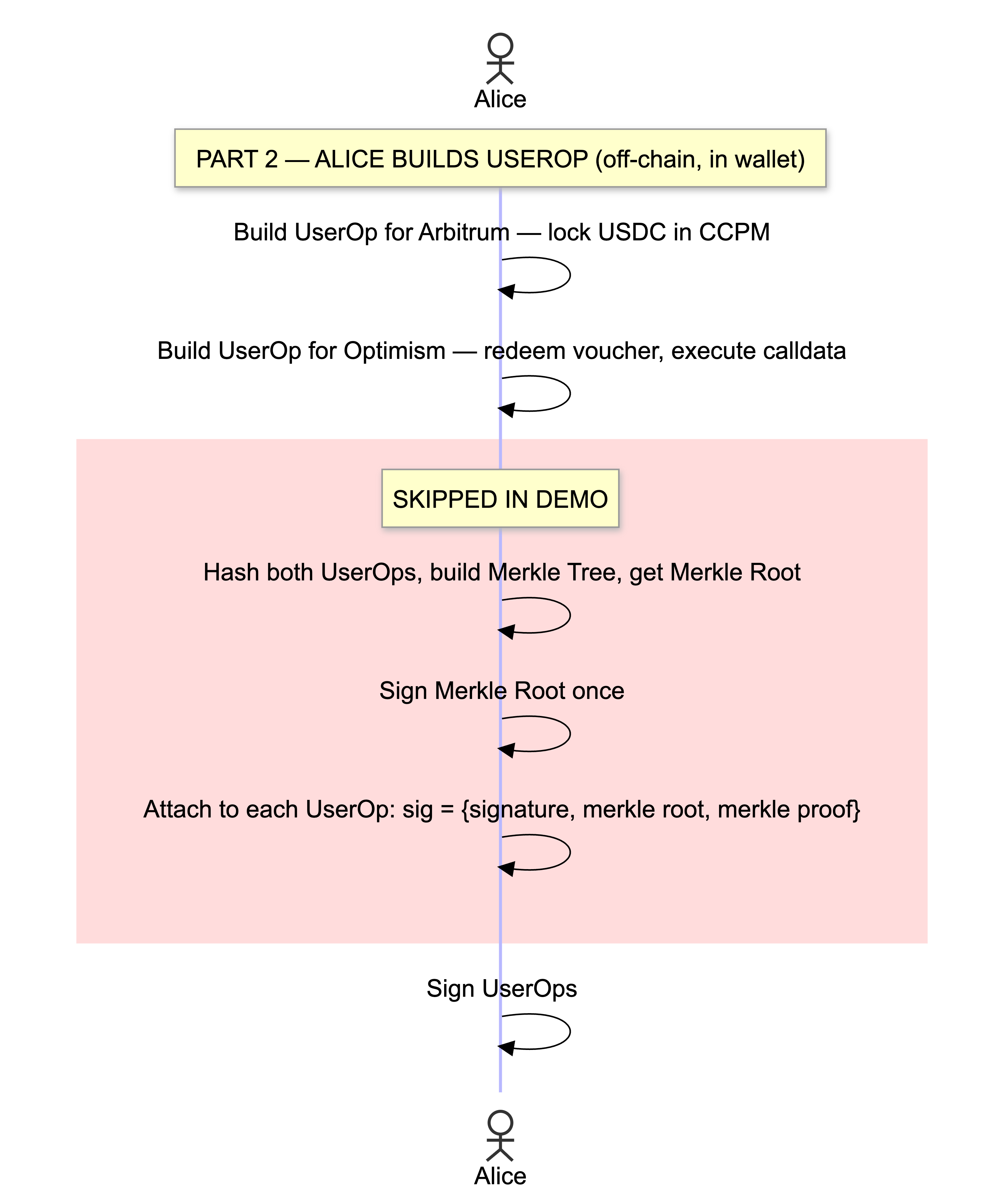}
\caption{Wallet-side construction of per-chain \texttt{UserOperation}s. The red block marks the one-signature Merkle aggregation path that was intentionally skipped in the current prototype; the in-figure label ``SKIPPED IN DEMO'' is preserved from the original workflow visualization.}
\label{fig:prep-demo-b}
\end{figure}

\section{Implementation Delta and Experimental Setup}

\subsection{Design Goals}
The framework is designed around three goals: reproducibility, protocol isolation, and engineering traceability. Reproducibility means that both L2 environments can be redeployed deterministically with the same contract topology and event flow. Protocol isolation means that the core voucher request and redemption path can be tested without requiring every production-grade component to be present. Engineering traceability means that validation failures can be localized to a specific step, such as signing, paymaster preprocessing, bundling, voucher issuance, or destination-chain execution.

\subsection{Prototype Scope and Delta from Intended EIL}
The intended EIL architecture includes multichain account abstraction, event-sourced orchestration, XLP-based voucher issuance, and dispute-aware settlement \cite{eil_overview}. Our prototype deliberately narrows that scope so the voucher path can be executed in a deterministic research environment. This difference should be explicit because the paper validates the executability of the core communication path, not the entire production EIL stack. Table~\ref{tab:delta} summarizes the main prototype deltas and their effect on validity.

\begin{table*}[!t]
\caption{Prototype realization versus intended EIL design}
\label{tab:delta}
\centering
\scriptsize
\renewcommand{\arraystretch}{1.15}
\begin{tabular}{|C{0.15\textwidth}|C{0.20\textwidth}|C{0.25\textwidth}|L{0.29\textwidth}|}
\hline
\textbf{Intended feature} & \textbf{Prototype realization} & \textbf{Rationale} & \textbf{Impact on validity} \\
\hline
One-signature multichain authorization
& Per-chain signing through a local multichain wrapper
& Keep the workflow executable while isolating the validation gap
& Validates protocol executability, not the final user-signing UX. \\
\hline

Standard smart-account compatibility
& Controlled wrapper with local \texttt{DummyAccount}-based execution path
& The current SDK path is not reproducible under a standard \texttt{SimpleAccount}-style hash check
& Demonstrates the incompatibility and a local repair, not universal wallet support. \\
\hline

Production bridge and settlement path
& Local dual-L2 setup with mock bridge components
& Deterministic deployment and replayable event flow
& Sufficient for voucher-path validation, not for final settlement claims. \\
\hline

Full reclaim, dispute, and settlement support
& Explicitly out of scope in the current prototype
& Repo implementation prioritizes core voucher execution
& Claims are restricted to the request--voucher--redeem lifecycle. \\
\hline
\end{tabular}
\end{table*}

\subsection{System Components}
Our testbed uses two local development chains representing Arbitrum-like and Optimism-like environments. On top of these chains, we deploy the paymaster, bridge-related, and destination validation components required by the EIL workflow. We then add mock bundlers that accept locally constructed \texttt{UserOperation} objects and event-driven XLP providers that listen for origin-chain voucher requests and return signed vouchers for destination-chain execution. This setup directly targets the Ethereum-side engineering problem studied in this paper: whether a voucher-based L2 interoperability path can be made executable and independently verifiable under realistic ERC-4337-style validation assumptions.

The use of mock bundlers is intentional. It enables controlled inspection of the exact object passed into validation and removes unrelated network-level variability from the experiment. Similarly, event-driven XLP providers allow the voucher issuance step to remain faithful to the architecture while still making the entire pipeline deterministic and locally reproducible. Table~\ref{tab:setup} summarizes the deployed environment, and Algorithm~\ref{alg:flow} lists the executable validation sequence. To support artifact availability and independent inspection, the prototype implementation used in this study is publicly available in the EIL\_research repository \cite{eil_repo}. Algorithm~\ref{alg:flow} summarizes the executable voucher workflow corresponding to the preparation stages in Figs.~\ref{fig:prep-demo-a} and~\ref{fig:prep-demo-b} and the public execution traces in Figs.~\ref{fig:sepolia-exec-a} and~\ref{fig:sepolia-exec-b}.

This local harness was also necessary because the tested SDK version embeds several assumptions that are not obvious from the public-facing configuration examples. In particular, chain-specific addresses are effectively resolved through \texttt{chainInfos[]} rather than the higher-level \texttt{deployments} field, the workflow depends on a custom \texttt{EntryPoint} address rather than a broadly supported public ERC-4337 deployment, and parts of the deployment metadata reference private Tenderly-backed RPC infrastructure. These assumptions do not negate the architectural idea, but they do mean that clean-room reproduction cannot rely on default SDK configuration or public bundler services alone. Here, standards-aligned refers to validation semantics---especially hash consistency between signed and later-validated \texttt{UserOperation} content---rather than to immediate out-of-the-box interoperability with public bundler ecosystems.

\begin{table}[!ht]
\caption{Experimental environment and deployed artifacts}
\label{tab:setup}
\centering
\scriptsize
\renewcommand{\arraystretch}{1.15}
\begin{tabular}{|C{0.24\columnwidth}|L{0.64\columnwidth}|}
\hline
\textbf{Item} & \textbf{Configuration} \\
\hline
Local chains
& Arbitrum-like Anvil chain (\texttt{chainId}=42161, port 8501) and Optimism-like Anvil chain (\texttt{chainId}=10, port 8503) \\
\hline

SDK packages
& \texttt{@eil-protocol/sdk} 0.1.2 and \texttt{@eil-protocol/accounts} 0.1.2 \\
\hline

Core contracts
& \texttt{EntryPoint}, \texttt{CrossChainPaymaster}, \texttt{SimpleAccountFactory}, \texttt{MockL2Bridge}, and \texttt{TestERC20} tokens \\
\hline

Local services
& Mock bundlers on ports 3000 and 3001, plus an event-driven XLP provider \\
\hline

Determinism support
& Create2-based deterministic contract deployment and fixed local service topology \\
\hline

Primary evidence
& Bundler logs, voucher-related events, destination execution logs, and final token-balance changes \\
\hline
\end{tabular}
\end{table}

\begin{algorithm*}[!t]
\caption{Validated voucher workflow in the current prototype}
\label{alg:flow}
\begin{algorithmic}[1]
\STATE Prepare the two L2 environments and register the XLP with the required liquidity on the origin and destination sides
\STATE Construct the per-chain \texttt{UserOperation} objects on the wallet side for the origin request and destination redemption paths
\STATE In the current prototype, skip the one-signature Merkle aggregation path and instead sign the per-chain \texttt{UserOperation} objects through the compatible smart-account wrapper
\STATE Submit the origin-chain batch, which transfers the selected amount and appends \texttt{addVoucherRequest(...)} for the destination chain
\STATE Let the origin-side \texttt{VoucherRequestCreated} event trigger the XLP provider to sign the voucher and call \texttt{issueVouchers(...)}
\STATE Attach the issued voucher to the destination-chain batch containing \texttt{useVoucher(...)} and the destination-side transfer action
\STATE Submit the destination-chain batch for voucher verification and execution
\STATE Verify the emitted evidence, including origin submission, voucher issuance, destination redemption, and the resulting destination-side asset release
\end{algorithmic}
\end{algorithm*}

These system components define a deliberately scoped research harness rather than a full production interoperability stack. The framework is not intended to emulate every aspect of a production network. Instead, it targets the minimal end-to-end path needed to make the protocol testable in a credible research setting. This includes deterministic deployment, repeatable event ordering, and explicit capture of evidence at each stage. This narrower scope is intentional because it allows the evaluation to focus on the verification gap and its repair rather than on infrastructure variability. In this way, the study emphasizes execution validity, reproducibility, and engineering traceability over broader performance claims.

\section{Failure Reproduction and Compatibility Repair}

\subsection{Failure Reproduction}
In the evaluated execution path, a user signs a \texttt{UserOperation}, but later paymaster-related bytes are no longer identical to the bytes observed by the local bundler-side validation flow. In the controlled harness, this creates a hash-consistency problem: the object used for local validation is not transparently identical to the representation treated as user-authorized by the prototype. The result is a reproducibility failure, because the voucher lifecycle cannot be exercised through the intended local sequence without an explicit compatibility layer.

This result should be interpreted narrowly. The prototype uses a controlled wrapper and a \texttt{DummyAccount}-based local execution flow to isolate the evaluated compatibility issue. It therefore does \emph{not} demonstrate that a production \texttt{SimpleAccount} path, every version of the EntryPoint, or every public bundler fails in the same way. It also does not claim that the observed behavior is a canonical EIL-wide defect. The package versions, local services, and contract topology used in the study are stated in Table~\ref{tab:setup}.

\begin{figure*}[!t]
\centering
\includegraphics[width=1\textwidth]{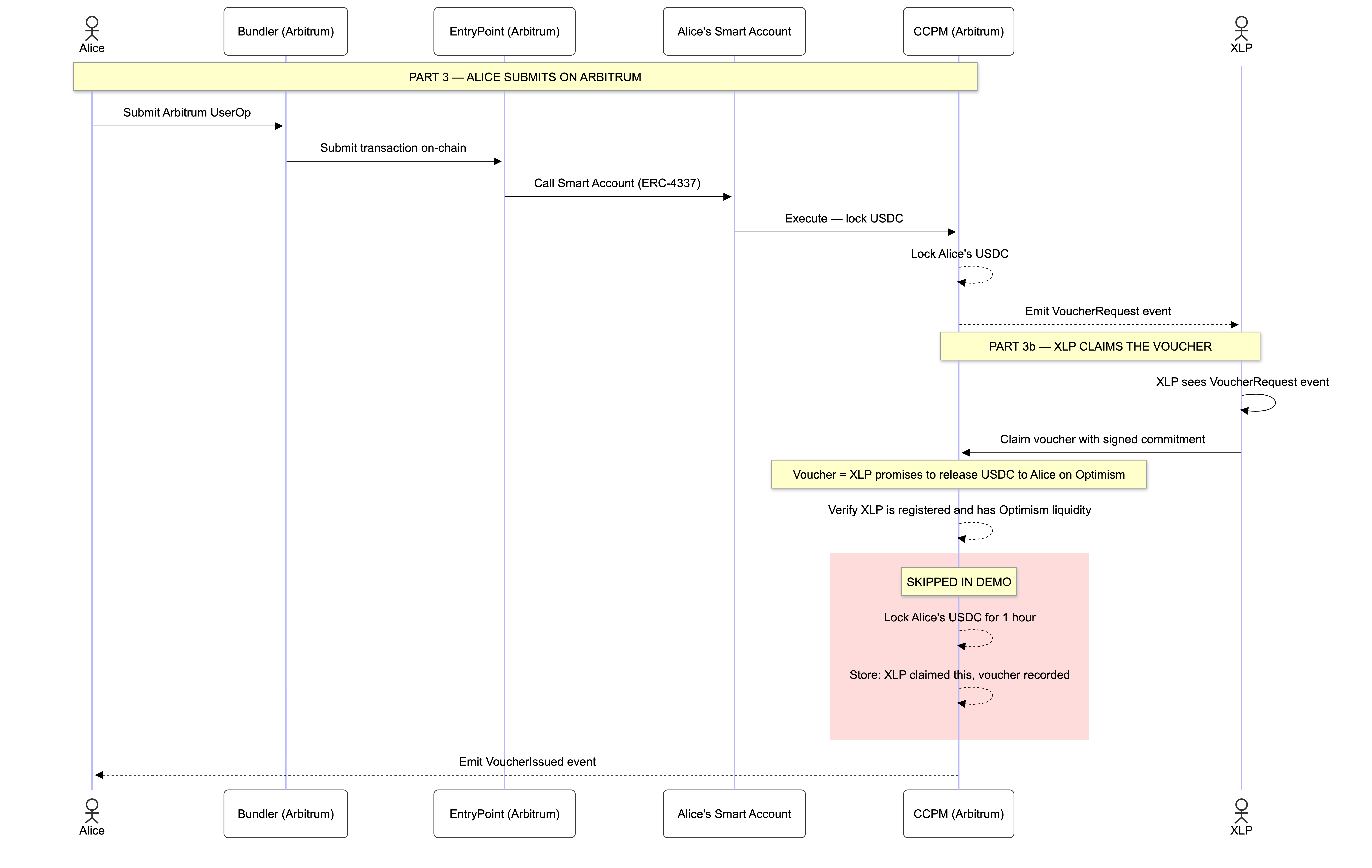}
\caption{Origin-chain submission on Arbitrum Sepolia, where Alice locks USDC, emits the voucher request, and enables the XLP-side claim/issuance path.}
\label{fig:sepolia-exec-a}
\end{figure*}

\subsection{Controlled Compatibility Mitigation}
The mitigation makes the local validation basis explicit. Rather than silently treating post-signature paymaster mutation as harmless, the local bundler normalizes the affected paymaster-data representation before it recomputes the harness validation hash. Concretely, it removes a paymaster-signature trailer identified by a fixed magic prefix and then evaluates the sanitized representation. Equations~\eqref{eq:strip} and~\eqref{eq:hsan} define the operation used by the prototype.

\begin{equation}
\tilde{p}=\mathrm{stripTrailer}(\texttt{paymasterData})
\label{eq:strip}
\end{equation}

\begin{equation}
H_{\mathrm{san}}(op)=H\bigl(op[\texttt{paymasterData}\leftarrow \tilde{p}]\bigr)
\label{eq:hsan}
\end{equation}

This is a prototype-level compatibility mitigation, not a claim that arbitrary byte stripping is safe. Its validity depends on the removed trailer being external to the authorization-relevant representation used by the controlled experiment. The implementation must therefore record the evaluated SDK and EntryPoint versions, the fixed prefix, and the exact byte-range handling in the released artifact. If the trailer corresponds to a versioned ERC-4337 paymaster-signature convention, the intended behavior must be checked against that version's hashing rules; otherwise, the mitigation must not be generalized beyond the local harness.

The mitigation is paired with a compatible multichain account wrapper that signs per-chain \texttt{UserOperation} objects and coordinates origin and destination batches needed for voucher issuance and redemption. It is not the final one-signature multichain solution, and the \texttt{DummyAccount}-based path is not evidence of universal wallet support. Instead, the wrapper provides a controlled, inspectable execution path for isolating the observed compatibility issue and restoring testability of the voucher lifecycle.

\subsection{Security and Scope Boundary}
The threat model is intentionally limited. The experiment does not evaluate malicious bundlers, malicious XLP providers, forged vouchers, production paymaster economics, L1 settlement attacks, dispute resolution, slashing, or bridge-security guarantees. It evaluates whether the local request--voucher--redeem sequence can be reproduced once the observed incompatibility in the evaluated SDK configuration is made explicit.

The sanitizer does not alter \texttt{sender}, \texttt{nonce}, \texttt{callData}, gas fields, or the stable paymaster-data prefix used by the prototype. Its purpose is solely to define a stable local validation representation after the identified trailer handling. This boundary is important: the experiment demonstrates controlled execution traceability, not a proof that the mitigation preserves all production authorization or security properties.

\subsection{Validated Execution Sequence}
After the repair is applied, the prototype executes the key workflow steps in order. A user-originated request is formed and signed on the origin chain, captured by the local bundler, emitted as a voucher-related event, observed by the XLP provider, transformed into a signed voucher, and then redeemed on the destination chain. The destination chain verifies the voucher and releases the corresponding asset or execution effect. This sequence is sufficient to validate the core communication and execution path claimed by the architecture, while remaining explicit about the prototype's reduced scope. Table~\ref{tab:gap} summarizes the failure mode, repair, and resulting executable path.

\begin{table*}[!t]
\caption{Verification gap and repair summary}
\label{tab:gap}
\centering
\small
\renewcommand{\arraystretch}{1.15}
\begin{tabular}{|C{0.12\textwidth}|L{0.34\textwidth}|L{0.22\textwidth}|L{0.22\textwidth}|}
\hline
\textbf{Stage} & \textbf{Verification gap} & \textbf{Repair} & \textbf{Outcome} \\
\hline
Signing
& The user signs stable \texttt{UserOperation} fields, but later paymaster-related data differs from the signed content.
& Normalize the affected validation input.
& Stable validation basis. \\
\hline

Bundling
& The bundler forwards an operation whose validation view contains mutated fields.
& Recompute the hash over the sanitized object.
& Reproducible local execution. \\
\hline

Destination redemption
& A valid origin-side request cannot be exercised downstream because the upstream validation failure blocks voucher redemption testing.
& Restore the end-to-end path with a compatible wrapper.
& Voucher lifecycle becomes testable. \\
\hline
\end{tabular}
\end{table*}

\begin{table*}[!t]
\caption{Representative evidence collected from the repaired local workflow}
\label{tab:evidence}
\centering
\scriptsize
\renewcommand{\arraystretch}{1.15}
\begin{tabular}{|C{0.12\textwidth}|C{0.12\textwidth}|C{0.32\textwidth}|L{0.32\textwidth}|}
\hline
\textbf{Evidence item} & \textbf{Prototype source} & \textbf{Representative value} & \textbf{Interpretation} \\
\hline
Two-L2 test environment
& Deployment and prototype setup
& Arbitrum-like localhost:8501 and Optimism-like localhost:8503 with deterministic contract deployment
& The validation target is reproducible and replayable across runs. \\
\hline

Initial user balances
& Prototype execution log
& Arbitrum: 20000 USDC; Optimism: 0 USDC
& The user initially holds source-chain assets only. \\
\hline

Configured transfer amount
& Cross-chain execution script
& \texttt{amtToBridge = 300000} base units = 0.3 USDC
& The destination-side balance change can be checked against a concrete configured amount. \\
\hline

Origin execution evidence
& Bundler / prototype log
& \texttt{Batch 0: done (tx: 0x41190e7f985f7923...)}
& The origin-chain \texttt{UserOperation} is accepted and mined. \\
\hline

Voucher issuance evidence
& XLP provider path
& \texttt{VoucherRequestCreated} triggers \texttt{issueVouchers(...)}; prototype log shows \texttt{Batch 1: voucherIssued}
& The off-chain XLP path reacts to the request and produces a signed voucher. \\
\hline

Destination execution evidence
& Bundler / prototype log
& \texttt{Batch 1: done (tx: 0xa8fc959fd822b00b...)}
& The destination chain verifies the voucher and executes the redemption batch. \\
\hline

Final asset evidence
& Prototype execution log
& Arbitrum: 19999.7 USDC; Optimism: 0.3 USDC
& A non-zero destination balance confirms that the voucher-based redemption path was actually exercised. \\
\hline
\end{tabular}
\end{table*}

\begin{table*}[!t]
\caption{Reviewer-verifiable public Sepolia transactions for the repaired voucher workflow}
\label{tab:sepolia-public-artifacts}
\centering
\small
\renewcommand{\arraystretch}{1.15}
\begin{tabular}{|C{0.17\textwidth}|C{0.22\textwidth}|L{0.51\textwidth}|}
\hline
\textbf{Network} & \textbf{Public Link} & \textbf{Evidence} \\
\hline
Arbitrum Sepolia
& \href{https://sepolia.arbiscan.io/tx/0x641b82211ccecee882e5daa2b12b33378edbb6f850c4387f18b81c042d9de1f7}{View origin transaction}
& Successful \texttt{handleOps} execution; \texttt{UserOperationEvent} records zero paymaster; ERC-20 logs show a 2~USDC origin-side transfer. \\
\hline

OP Sepolia
& \href{https://sepolia-optimism.etherscan.io/tx/0xfbf13ef74b7019797189754167b12b2948d9ff40d391321e41a05aae183fa9f7}{View destination transaction}
& Successful \texttt{handleOps} execution; \texttt{UserOperationEvent} records a nonzero paymaster; ERC-20 logs show a 2~USDC destination-side transfer. \\
\hline
\end{tabular}
\end{table*}

\begin{figure*}[!t]
\centering
\includegraphics[width=1\textwidth]{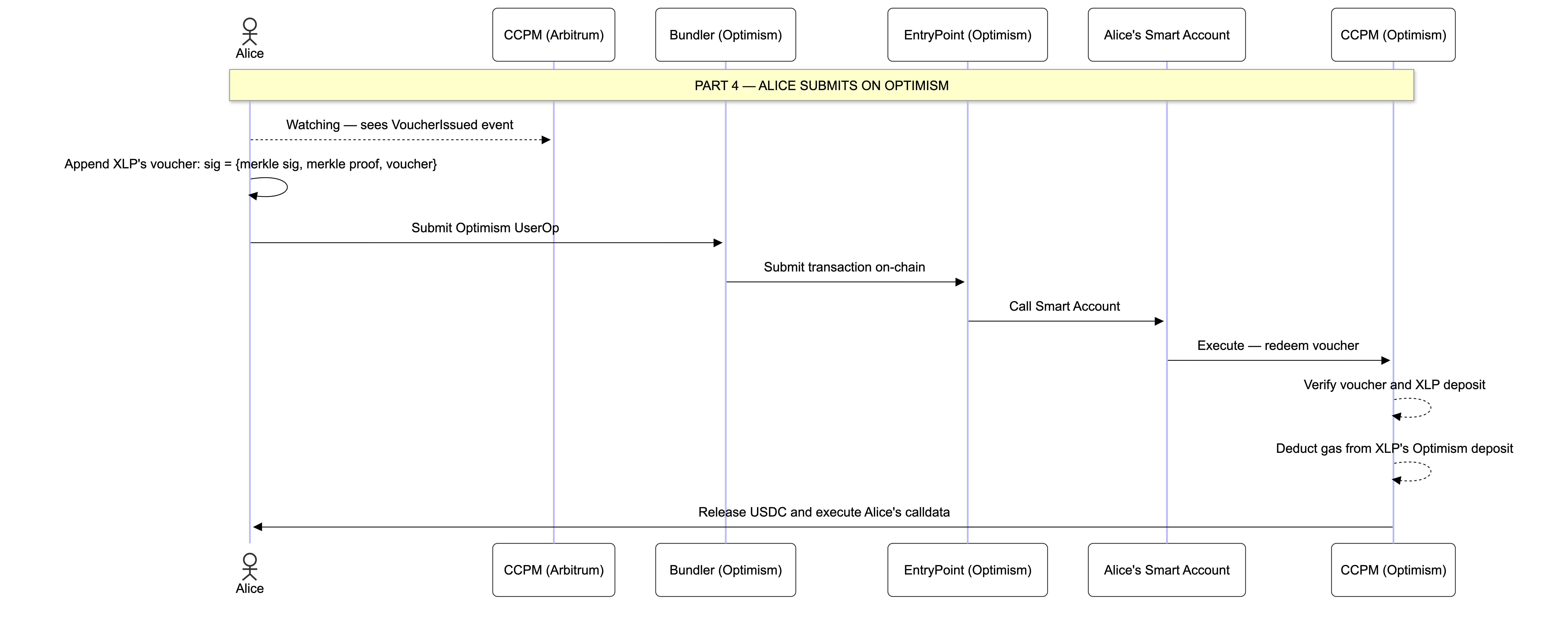}
\caption{Destination-chain submission on OP Sepolia, where the issued voucher is appended, verified, and redeemed to release the destination-side USDC.}
\label{fig:sepolia-exec-b}
\end{figure*}

\section{Experimental Validation and Discussion}

\subsection{Validation Questions}
Our evaluation is centered on controlled execution validity and traceability rather than broad performance, security, or generality claims. We ask two questions: (1) can the evaluated SDK configuration be reproduced sufficiently to isolate the paymaster-data compatibility issue, and (2) can the prototype-level mitigation restore an inspectable end-to-end request--voucher--redeem execution path in a dual-L2 environment? We therefore record evidence for request creation, bundler acceptance, voucher-event observation, off-chain voucher issuance, destination-chain verification, and final execution. Table~\ref{tab:evidence} reports representative evidence from one repaired local workflow.

Deterministic deployment makes the testbed suitable for replay, but the current paper does not report a repeated-run, cross-machine, or state-root comparison study. Accordingly, the evidence should be read as execution traceability for a controlled workflow rather than a quantitative claim of deterministic equivalence across machines or software versions. A future artifact evaluation should report repeated fresh deployments, per-stage success rates, exact package and contract commits, and comparable event-sequence or state-root summaries.

\subsection{Public Sepolia Testnet Verification}
To complement the controlled local validation above, we additionally executed a public testnet demonstration on Arbitrum Sepolia and OP Sepolia. This run serves as reviewer-verifiable corroboration of the repaired request--voucher--redeem path, rather than as a replacement for the deterministic Anvil-based experiment summarized in Table~\ref{tab:evidence}. The setup stages are illustrated in Figs.~\ref{fig:prep-demo-a} and~\ref{fig:prep-demo-b}, and the corresponding public execution traces are shown in Figs.~\ref{fig:sepolia-exec-a} and~\ref{fig:sepolia-exec-b}. The main public artifacts are summarized in Table~\ref{tab:sepolia-public-artifacts}. This public corroboration run used a different demonstration amount (2~USDC rather than the 0.3~USDC local test amount in Table~\ref{tab:evidence}), but it preserves the same request--voucher--redeem logic and therefore serves the same validation purpose.

The public traces are consistent with the repaired workflow in Algorithm~\ref{alg:flow}. On Arbitrum Sepolia, the origin-side \texttt{handleOps} execution records a zero paymaster, matching the current prototype configuration, and the ERC-20 logs confirm a 2~USDC origin-side transfer. On OP Sepolia, the destination-side \texttt{handleOps} execution records a nonzero paymaster, and the ERC-20 logs confirm a corresponding 2~USDC destination-side transfer. Auxiliary explorer-visible artifacts, including the common \texttt{EntryPoint} and the verified OP-side \texttt{CrossChainPaymaster}, remain available through the linked transactions but are omitted from the main table for compactness. Together, these public artifacts provide directly inspectable evidence for origin submission and destination redemption under the repaired voucher workflow.

The public artifacts show that the controlled request--voucher--redeem workflow can be inspected at the points that matter for the prototype: request construction, bundler acceptance, event-triggered voucher issuance, destination-side verification, and final state transition. They do not by themselves establish production compatibility or resolve the broader security questions of bridge settlement and account authorization. The contribution is therefore a bounded validation methodology that helps separate protocol intent from SDK-specific execution artifacts and addresses a recurring reproducibility challenge in complex blockchain systems \cite{verification_gap_blockchain}.

The resulting framework is still a research prototype, not a full production implementation. Even so, it lowers the cost of comparing interoperability designs, testing wallet integrations, and diagnosing where cross-rollup workflows fail. More broadly, the study suggests that future Ethereum interoperability work should pair new abstractions with validation-ready toolchains that wallets, infrastructure providers, and researchers can independently reproduce.\FloatBarrier
\section{Discussion and Scope}
The current prototype validates a narrowly defined voucher-based execution path and cross-L2 communication logic in a controlled dual-L2 environment. It isolates the evaluated compatibility issue by using local services, a compatible account wrapper, and a \texttt{DummyAccount}-based execution path. This design enables engineering traceability, but it also limits the claim: the prototype does not establish compatibility with a production \texttt{SimpleAccount}, a public ERC-4337 bundler ecosystem, or the canonical implementation of every EIL component.

The work should therefore be understood as an evaluation of an evolving SDK configuration rather than a general statement about the EIL protocol. Its main value is to make the implementation boundary visible: which components are executable today, which assumptions are embedded in the local harness, and which claims require further upstream or standard-account validation. The next technical step is to test an upstream-corrected path, if available, with an explicitly versioned EntryPoint, a standard smart account, and independently reproducible public-bundler execution.

\section{Conclusion}
This paper presents a reproducible compatibility study of voucher-based L2 interoperability in an evaluated EIL SDK configuration. We show that post-signature paymaster-data handling can obstruct clean-room reproduction when the local validation representation diverges from the representation treated as user-authorized by the prototype. We then provide a controlled mitigation and a two-L2 validation harness that makes the request--voucher--redeem path executable and inspectable.

The contribution is threefold:
\begin{enumerate}
\item \textbf{Controlled validation framework:} a two-L2 environment that captures evidence across request creation, voucher issuance, destination redemption, and asset release.
\item \textbf{Bounded compatibility mitigation:} an explicit sanitizer and wrapper for the evaluated local execution path, with no claim of universal ERC-4337 or wallet compatibility.
\item \textbf{Inspectable systems artifact:} a reproducible basis for separating protocol intent from SDK-specific execution behavior in voucher-based interoperability experiments.
\end{enumerate}

The remaining work is equally important: validate the path with a standard smart account and versioned EntryPoint semantics, document upstream issue and patch status where applicable, evaluate repeated-run determinism, and extend the scope to one-signature multichain authorization, settlement, disputes, and production security assumptions.
\bibliographystyle{IEEEtran}
\bibliography{references}

\end{document}